\documentclass[letterpaper,11pt]{article}
\usepackage{bbm}
\usepackage{latexsym,amssymb,amsmath, bm}

\usepackage{amsthm}
\usepackage{lipsum,graphicx,multicol}
\usepackage{threeparttable}
\usepackage{tabularx}
\usepackage{longtable}
\usepackage{array}
\usepackage[colorlinks=true,linkcolor = red,citecolor = blue]{hyperref}
\usepackage{setspace,color}
\usepackage[sort&compress]{natbib}
\usepackage{graphicx}
\usepackage[table]{xcolor}
\usepackage{float}
\usepackage{authblk}
\usepackage{leftidx}
\usepackage{geometry}
\usepackage[title,titletoc]{appendix}
\usepackage{epstopdf}
\usepackage{actuarialangle}
\usepackage{scrextend}
\usepackage{booktabs}
\usepackage{lscape}
\usepackage{caption}
\usepackage{subcaption}
\usepackage{indentfirst}
\usepackage{float}
\usepackage{adjustbox}
\usepackage[linesnumbered,lined,boxed,commentsnumbered]{algorithm2e}
\allowdisplaybreaks

\usepackage{enumitem}
\setlist{
  listparindent=\parindent,
  parsep=0pt,
}
\usepackage[title]{appendix}
\usepackage{multirow,bigdelim}
\usepackage{bbm}
\usepackage{tikz}
\usepackage{adjustbox}
\usetikzlibrary{shapes,arrows,positioning,fit,calc}

\usepackage{bm}

\usepackage{url}
\usepackage{breakurl}

\usetikzlibrary{decorations.pathmorphing}

\DeclareFontShape{OMX}{cmex}{m}{b}{<-> cmexb10}{}
\SetSymbolFont{largesymbols}{bold}{OMX}{cmex}{m}{b}
\usepackage{tcolorbox}
\usepackage{amsmath,bm}
\usepackage{arydshln}
\usepackage{mathtools}
\usepackage{mhchem}
\usepackage{gensymb}

\usepackage{setspace}
\usepackage{accents}


\usepackage[table]{xcolor}

\definecolor{darkread}{rgb}{0.7, 0, 0}
\definecolor{darkbrown}{rgb}{0.55, 0.2, 0.15}
\definecolor{darkblue}{rgb}{0.1,0.1,0.6}
\definecolor{darkgreen}{rgb}{0.1,0.5,0.2}

\begin{document}

\title{\Large\textbf{Robo-Advisors Beyond Automation: Principles and Roadmap for AI-Driven Financial Planning}}

\author{
\qquad {Runhuan Feng}\thanks{School of Economics and Management, Tsinghua University, Beijing, China.
Email: \href{mailto:}{\color{magenta} fengrh@sem.tsinghua.edu.cn}}\;\thanks{National Center for Economic Research, Tsinghua University, Beijing, China.} \qquad
\qquad {Hong Li}\thanks{Department of Economics and Finance, Gordon S. Lang School of Business and Economics, University of Guelph, Guelph, Canada. Email: \href{mailto:lihong@uoguelph.ca}{\color{magenta}lihong@uoguelph.ca} } \qquad
\qquad {Ming Liu}\thanks{Corresponding Author. School of Applied Economics, Guizhou University of Finance and Economics, Guiyang, Guizhou, China. Email: \href{mailto: ming_liu1@126.com}{\color{magenta} ming\underline{~}liu1@126.com} }
\qquad
}

\date{\today}

\maketitle

\begin{abstract}
Artificial intelligence (AI) is transforming financial planning by expanding access, lowering costs, and enabling dynamic, data-driven advice. Yet without clear safeguards, digital platforms risk reproducing longstanding market inefficiencies such as information asymmetry, misaligned incentives, and systemic fragility. This paper develops a framework for responsible AI in financial planning, anchored in five principles: fiduciary duty, adaptive personalization, technical robustness, ethical and fairness constraints, and auditability. We illustrate these risks and opportunities through case studies, and extend the framework into a five-level roadmap of AI financial intermediaries. By linking technological design to economic theory, we show how AI can either amplify vulnerabilities or create more resilient, trustworthy forms of financial intermediation.
\end{abstract}




\section{Introduction}
\label{sec:introduction}

In financial markets, decision-making is often constrained by information asymmetry, where one party holds superior knowledge about products, risks, or prices, creating adverse selection and moral hazard problems \citep{akerlof1978market}. At the same time, transaction frictions—including search and monitoring costs, contractual rigidities, and limited market transparency—raise barriers to efficient exchanges and increase reliance on intermediaries \citep{coase1993nature, meckling1976theory, jensen2019theory}. These structural challenges motivate the enduring role of financial intermediation.

Financial institutions long relied on human advisors to reduce information asymmetries, mitigate transaction frictions, and help individuals and firms navigate complex financial decisions. Traditional advisory models, however, are fraught with inefficiencies: incentive misalignments through commission-driven sales \citep{edwards2017conflicts}, subjective biases in judgment \citep{sahi2017psychological}, and limited scalability that constrains access to professional advice \citep{howcroft2003banker, morgan1994commitment, vargo2004evolving, hunt2011determinants, soderberg2013relationships}. Over the past two decades, advances in artificial intelligence (AI) and digital finance have promised to transform this landscape by automating advisory processes, reducing costs, and extending financial planning to underserved populations \citep{belanche2019artificial, belanche2020service, adam2019investment, morana2020effect}. 

The rise of robo-advisors epitomizes this technological shift. Platforms such as Wealthfront, Betterment, and Vanguard Personal Advisor Services have demonstrated the potential for algorithmic tools to democratize access to investment advice, improve efficiency through automation, and enforce discipline in portfolio rebalancing \citep{wexler2021robo, ashrafuzzaman2025ai}. Yet, adjacent digital brokerages and social-trading platforms—\emph{distinct from robo-advisors} because they center on self-directed trading and copy-trading rather than automated portfolio management—expose a different set of risks that can spill over into advisory contexts. Robinhood’s reliance on gamified interfaces and payment for order flow illustrates the dangers of behavioral manipulation and conflicted incentives \citep{tierney2022investment}, while eToro’s opaque ranking and copy-trading mechanisms reveal risks of algorithmic opacity, miscommunicated risk, and systemic fragility \citep{gregersen2024platform}. In both cases, digital innovations lower entry barriers while creating new channels for client harm. These tensions highlight that AI-enabled finance, far from resolving long-standing inefficiencies, may reconfigure them in harder-to-detect forms (see Section~\ref{sec:casestudy}).

Despite rapid adoption, scholarly and regulatory frameworks for classifying and governing AI in financial advice remain underdeveloped. Much of the existing literature evaluates robo-advisors through the lens of technological maturity—contrasting rule-based calculators with more autonomous, self-learning systems \citep{arrieta2020explainable, huang2025survey, oelschlager2024evaluating}. Others emphasize regulatory or ethical dimensions, such as explainability, bias, and consumer protection \citep{zhu2024implementing, inavolu2024exploring, challoumis2024future}. What remains missing is an integrated framework that situates AI systems not only along a technological continuum but also in relation to the economic functions they are meant to perform: reducing information asymmetry, mitigating adverse selection, and aligning incentives to overcome moral hazard. Without such a perspective, assessments risk overlooking whether technical advances actually improve market functioning or merely shift inefficiencies into new forms.

This paper contributes to closing that gap in three ways. First, through case studies of Robinhood, eToro, and contemporary AI applications, we illustrate how design choices in digital platforms can replicate or exacerbate vulnerabilities in financial intermediation, grounding the analysis in real-world practices. Second, we advance a normative framework of five foundational principles—fiduciary duty, adaptive personalization, technical robustness and resilience, ethical and fairness constraints, and auditability and accountability—that specify the conditions under which AI-based financial planning can align with client welfare and systemic stability. These principles are not claimed by prior studies; rather, they synthesize themes that have been discussed separately in the literatures on financial advice and AI ethics. For instance, research on professional responsibility and integrated planning highlights fiduciary concerns \citep{irving2012financial, chieffe1999integrated}; work on digital financial services and robo-advisory emphasizes personalization and client alignment \citep{belanche2019artificial, adam2019investment, morana2020effect}; and studies in explainable and trustworthy AI stress robustness, fairness, and accountability \citep{arrieta2020explainable}. Our contribution is to integrate these strands into a coherent framework tailored to AI-enabled advisory services.
Third, we propose a development roadmap that categorizes AI intermediaries into five maturity levels, from basic calculators to anticipatory “super‑intelligent” planners. Unlike prior taxonomies, this roadmap explicitly links maturity levels to the mitigation of economic inefficiencies, offering a functional assessment of progress and limitations.

By combining empirical cases, normative principles, and a developmental roadmap, the paper advances both scholarly and practical debates on AI in finance. For scholars, it integrates insights from financial economics and AI governance, bridging literatures that often remain siloed. For practitioners and regulators, it provides actionable criteria to evaluate whether specific AI systems enhance or undermine financial advice. Ultimately, the analysis demonstrates that efficiency and scalability alone are insufficient: without careful attention to fiduciary alignment, fairness, robustness, and accountability, AI may extend the reach of financial advice but fail to secure its legitimacy. {The concepts of \emph{information asymmetry} and \emph{transaction frictions} introduced above serve as benchmarks throughout the paper for judging whether AI-based advice reduces—or merely relocates—market imperfections.}

The remainder of the paper proceeds as follows. Section~\ref{sec:background} examines the opportunities and limitations of robo-advisors. Section~\ref{sec:casestudy} develops case studies. Section~\ref{sec:Prin} outlines five foundational principles for responsible AI-driven financial planning. Section~\ref{sec:roadmap} presents a development roadmap linking levels of AI maturity to the mitigation of financial inefficiencies. Section~\ref{sec:conclusion} concludes.

\section{From Human to Robo-Advisors}
\label{sec:background}
The rise of robo-advisors must be understood against the backdrop of how financial advising has evolved as a response to persistent market imperfections. This section traces that evolution: it begins with the role of human financial advisors, examines the inefficiencies that limit their effectiveness, and then turns to the emergence of algorithmic advisory systems as a potential remedy.

\subsection{The Emergence of Financial Advisors}

Financial planning developed in response to what economists describe as the persistent imperfections of financial markets. These are not momentary anomalies, but enduring conditions—borrowers and investors rarely hold the same information, contracts are costly to write and enforce, and individual choices are often shaped by bounded rationality. Such features, well documented in the economics of information and transaction cost theory \citep{akerlof1978market, coase1993nature, langlois1998transaction}, explain why financial markets never reach the frictionless ideal. Intermediaries such as banks, insurers, and asset managers arose to soften these imperfections by pooling resources and reducing information gaps \citep{diamond1986banking}. Yet solutions often came with their own complications: product designs grew opaque, incentives became layered, and retail clients faced decisions they could not easily evaluate \citep{thaler1985mental}. A typical example is the difficulty investors face in assessing mutual fund fees, where layered charges and obscure expense ratios make products hard to compare for most retail clients \citep{fisch2014retail}. In this environment, professional advisors emerged to help clients interpret complexity, align strategies with goals, and safeguard their interests.

Although advisors operate in varied settings—banks, broker–dealers, insurers, and, more recently, digital platforms—their core functions converge: reducing information gaps, lowering search and compliance costs, and translating complex choices into implementable plans. Their contribution is particularly evident in addressing two enduring problems.

The first is \emph{information asymmetry}: most clients lack access to comprehensive financial data and the expertise required to interpret it. Advisors bridge this gap by consolidating dispersed market signals, product features, and regulatory developments into actionable strategies that reflect clients’ risk preferences, income stability, and long‑term objectives. The second is \emph{transaction frictions}: financial decisions impose search costs on clients and compliance costs on institutions \citep{benston1976transactions, morgan1994commitment, hunt1994relationship}. These frictions have intensified with the expansion of consumer protection regulation. Rules designed to prevent fraud and mis-selling—such as detailed disclosure requirements or “know-your-customer” obligations—undoubtedly strengthen investor safeguards, yet they also generate new layers of paperwork and procedural complexity. Prospectuses and fee statements, for instance, are meant to increase transparency, but in practice they overwhelm most retail clients with  technical language and length. Similarly, suitability tests and ongoing reporting create administrative burdens that raise costs for providers and slow the decision process for clients.
Advisors reduce these burdens by simplifying product access, cultivating long-term relationships, and ensuring that decisions conform to regulatory standards \citep{vargo2004evolving, hunt1994relationship}. In this way, they operate simultaneously as translators of information and as cost-reducing agents.

This historical evolution highlights the dual role of financial advisors: to simplify complexity and to reduce frictions in decision‑making. Yet it also foreshadows the vulnerabilities of the advisory model. As the next subsection discusses, the very structures that give advisors their value also create inefficiencies that limit their effectiveness and, in some cases, amplify the market failures they were meant to resolve.

\subsection{Inefficiencies of Traditional Financial Advisors} \label{sec:inefficiencies}

Although financial advisors were intended to mitigate market failures, in practice they often introduce inefficiencies of their own. The most pervasive issues are incentive misalignments and information asymmetries \citep{monti2014retail, golec1992empirical}, which manifest as moral hazard, adverse selection, and broader structural distortions. Unlike the market-level asymmetry between borrowers and lenders discussed above, this asymmetry arises within the advisor–client relationship: advisors typically possess superior knowledge of financial products, fee structures, and regulatory nuances, which gives them scope to steer clients toward choices that maximize commissions rather than client welfare. These are not minor operational frictions but systemic weaknesses that undermine client welfare and reduce allocative efficiency.

Moral hazard arises when advisors, once a transaction has taken place, exploit their informational advantage to serve their own interests rather than the client’s. Bank managers may recommend structured deposits with high commissions that poorly match client risk profiles \citep{edwards2017conflicts, reurink2019financial}. Securities advisors may encourage excessive trading (“churning”) or promote underwritten stocks to meet internal sales targets \citep{edesess2007big}. Insurance agents sometimes exaggerate benefits or obscure exclusions, leading clients to purchase costly or unsuitable coverage \citep{prince2017insurance}. In each case, superior product knowledge combined with sales-driven incentives creates strong pressures to distort advice.

Adverse selection represents a related but distinct problem. Risk-seeking or financially unsophisticated clients are disproportionately drawn to aggressive or opaque products, while more cautious clients often withdraw from advisory relationships. The result is a skewed client base that reduces the sustainability of financial services. High-yield bank products, for example, tend to attract risk-tolerant investors while deterring conservative ones \citep{hermalin1999risks, defond2016client}. Securities advisors frequently emphasize leverage, which alienates risk-averse households, and in life insurance markets healthier individuals often exit, leaving a pool concentrated with higher-risk clients. In each setting, advisors could amplify rather than correct adverse selection.

At the core of these patterns lies a principal--agent conflict. Compensation structures in many advisory channels reward short-term product sales rather than long-term client outcomes \citep{burke2015impacts}. Institutional practices add further layers of inefficiency: banks and insurers often privilege proprietary products, fragmenting markets and distorting allocation \citep{clark1976soundness}; and short-term performance pressures encourage sales targets and shortcuts in due diligence, with attendant reputational and legal risks \citep{kumar1999impact, dallas2011short}. Relational information imbalances compound these incentives by making it difficult for retail clients-often ill-equipped to evaluate advice-to verify quality \citep{akerlof1978market, arrow1978uncertainty}.

Even personalization—the central promise of advisory services—remains limited. Evidence from Canadian households \citep{foerster2017retail} shows that advisor characteristics, such as their own allocation preferences, explain more variation in client portfolios than client attributes such as risk tolerance, age, or investment horizon. In practice, this means that clients with very different profiles often end up with portfolios that look strikingly similar, reflecting the advisor’s imprint more than the client’s circumstances. It is in this sense that the advisory process tends toward a “one-size-fits-all” approach, undermining its promise of personalization. The costs are tangible: advised portfolios carry average annual expenses of about 2.5\%, roughly 1.5 percentage points higher than comparable lifecycle funds.

Taken together, these findings point to a paradox. Financial advisors exist to reduce information gaps, simplify choices, and guide long-term planning, yet in practice they often replicate the very frictions they are meant to resolve. Incentive structures push toward product sales rather than client outcomes; relational asymmetries leave clients dependent on advice they cannot easily verify; institutional loyalties to proprietary products fragment markets; short-term pressures weaken diligence; and the promise of personalization collapses into standardized portfolios. What should function as a corrective mechanism thus becomes another source of distortion, undermining both efficiency and trust.

It is in this sense that the recent rise of robo-advisors can be seen as a response to persistent weaknesses in the human model. By automating portfolio construction, standardizing compliance, and reducing conflicts of interest, digital platforms hold out the promise of more consistent alignment between advice and client welfare.

\subsection{Robo-Advisors: Promise and Limitations}
 
A robo-advisor—also referred to as a robo-advisory service—can be understood as a service robot built on an AI-empowered autonomous information system. Through interactive interfaces, it provides financial advisory services with minimal or no human intervention \citep{day2018ai, jung2018robo}. In practice, most platforms follow a broadly similar structure. Clients are first onboarded through questionnaires that capture financial goals, investment horizons, and risk tolerance. Portfolios are then constructed using frameworks such as Modern Portfolio Theory (MPT) \citep{markowitz1952modern, elton1997modern, black1990asset, black1992global} or its extensions. The system subsequently monitors market conditions, rebalances portfolios, and applies features such as tax-loss harvesting or dividend reinvestment. This model has expanded access to investment services by lowering costs, ensuring disciplined execution, and reaching segments of the population traditionally underserved by human advisors.

The integration of artificial intelligence (AI) has markedly reshaped the development trajectory of robo-advisors. Whereas early platforms relied primarily on standardized portfolio optimization models, contemporary systems increasingly embed advanced techniques such as machine learning, natural language processing, and recommendation algorithms \citep{vidler2024recommender, zatevakhina2019recommender}.  These tools allow for more nuanced forms of personalization and data-driven decision support. Chatbots and virtual assistants extend the availability of advisory services by responding to routine queries at any time \citep{kyrylenko2024chatbots}, while predictive analytics facilitate the continuous monitoring of market conditions and the timely adjustment of portfolio allocations \citep{schrettenbrunner2023artificial}. At the same time, recommendation systems adapt investment strategies by identifying behavioral patterns across users, and natural language processing contributes to regulatory compliance by transforming complex legal requirements into structured, machine-readable rules \citep{vidler2024recommender}. Collectively, these innovations indicate a transition of robo-advisors from narrow portfolio managers toward integrated platforms for holistic financial planning—delivering advice that is scalable, cost-efficient, and consistent in execution.

Within this transformation, an important distinction emerges between the roles of \textit{generative AI} and \textit{analytical AI}. Generative AI primarily enhances the communicative and interpretive dimensions of advisory services: it enables more natural interaction with clients, renders technical investment logic into accessible explanations, and generates scenario-based narratives that improve financial literacy and decision confidence. Analytical AI, by contrast, forms the quantitative backbone of robo-advisory operations. It drives the optimization of portfolio allocations, the detection of anomalies in market dynamics, the forecasting of portfolio risk, and the tailoring of investment strategies to client-specific constraints through statistical learning and predictive modeling. In combination, these two strands of AI perform complementary functions: generative AI strengthens trust and engagement by humanizing interactions, while analytical AI ensures rigor, discipline, and consistency in financial calculations. Their integration is therefore pivotal in reconciling the scalability of automated platforms with the normative expectation of client-centered financial planning.

{Before turning to their limitations, it is worth considering how robo-advisors seek to overcome the shortcomings of human advisors discussed in Section 2.2. Automation reduces the search and paperwork costs that often weigh heavily on clients, while digital interfaces lower entry barriers for households traditionally excluded from professional advice. By charging fees tied to assets under management rather than commissions on product sales, many platforms attempt to avoid the incentive misalignments that plagued traditional models. Algorithms also bring a degree of consistency, rather than relying on the subjective judgment of an individual advisor, portfolio recommendations are generated according to transparent, rule-based procedures. Finally, the use of client questionnaires and data analytics introduces a scalable form of personalization, ensuring that asset allocations can at least be adjusted to broad differences in risk tolerance and investment horizon.}

Despite these opportunities, robo-advisors face important constraints that prevent them from serving as a full substitute for human advisors. One such consequence is that robo-advisors focus primarily on investment-related advice and are largely limited to portfolio management, including asset allocation, tax optimization, and rebalancing. This is a result of their designs which are built to address the need for efficient, low-cost investment management rather than to provide comprehensive, long-term financial planning. Unlike human advisors, who offer personalized strategies that encompass retirement planning, estate management, and tax optimization across multiple life stages, robo-advisors are generally restricted to automated investment solutions that do not extend beyond client portfolios.

Given this largely investment-centric scope, automation does not by itself resolve the inefficiencies documented in Section~\ref{sec:background}. Two sets of concerns follow—first, issues that persist from human advisory models, and second, challenges introduced or amplified by AI-mediated design.

{The first limitation concerns incentive alignment, a problem robo-advisors inherit—albeit in new forms—from human advisors. Instead of commissions, conflicts arise from the economics of digital platforms themselves. Portfolios often privilege proprietary funds issued by the same institutions that operate the advisory service, creating an incentive to channel client assets into products that generate additional management fees. Business models based on order routing, securities lending, or revenue-sharing with affiliates can also bias recommendations in ways that are not visible to end users \citep{packin2024hooked}. Even the algorithms that structure client onboarding may be designed to upsell premium services or encourage product take-up beyond what is strictly optimal for investors. These mechanisms show that, although robo-advisors replace sales commissions with seemingly neutral algorithms, they do not eliminate the principal–agent tension between provider revenues and client outcomes.}

The {second} concerns behavioral manipulation and incentive alignment. Some platforms employ gamification strategies—such as push notifications, achievement badges, or simplified “swipe-to-trade” interfaces—that borrow from behavioral science and gaming mechanics to boost engagement \citep{zhu2024implementing}. While these design choices may make investing more accessible, they risk encouraging impulsive behavior, excessive trading, and poor risk assessment \citep{wendt2022regulation}. Business models based on payment for order flow or transaction volume compound the problem, as they create conflicts of interest that prioritize trading activity over client welfare and obscure the true cost of transactions \citep{, packin2024hooked}. In this sense, robo-advisors may replicate the same incentive distortions that compromised traditional financial advice, but in a digital form that can affect users at scale. 

A {third} limitation arises from issues of algorithmic opacity and the reliability of automated advices. Most robo-advisors rely on proprietary scoring models or recommendation engines whose internal workings are not visible to clients \citep{lourencco2020whose}. The lack of transparency means that investors cannot readily assess how recommendations are derived or whether these outputs genuinely reflect their stated objectives. Unlike certified human advisors who receive standardized professional training and are required to abide by professional code of conducts, Robo-advisors are currently developed independently by industry competitors without external auditing or third-party reviews. Moreover, existing research suggests that the advice generated by such systems is not always stable: clients with similar profiles may receive different recommendations depending on how the model is specified, how often the data are updated, or even how minor differences in survey responses are weighted \citep{kofman2025scoring}. From the user’s perspective, such inconsistencies are difficult to interpret, raising doubts about whether shifts in portfolio guidance are driven by meaningful market information or by arbitrary model adjustments. Regulators have echoed these concerns, noting that opaque machine-learning tools complicate both investor protection and supervisory oversight \citep{kothandapani2025ai}. In this respect, the central challenge lies less in the possibility of large-scale systemic disruption than in the inability of robo-advisors to provide transparent and consistently interpretable advice at the individual level.

Alongside these structural concerns, robo-advisors also face several more technical and behavioral limitations. Many platforms rely on similar portfolio models, leading to strategy convergence and collective vulnerability during crises \citep{liu2023judge}. Their standardized frameworks often fail to adapt to local market structures, as seen in the underperformance of robo-advisors in China’s A-share market \citep{guo2020regulating}. User skepticism remains a challenge, with surveys reporting widespread abandonment of platforms after losses \citep{benavent2021vanguard, bianchi2023robo}. Personalization is typically superficial, relying on short questionnaires that classify clients into broad risk categories \citep{adam2019investment}. Finally, financial planning is inherently dynamic, yet most robo-advisors apply relatively static rules, leaving them poorly suited to address evolving life-cycle needs \citep{zhu2024implementing}.

In sum, robo-advisors embody both the promise and the constraints of automating financial advice. Their efficiency, accessibility, and disciplined execution represent genuine progress, but their behavioral design choices, opaque algorithms, and systemic vulnerabilities risk reproducing—and in some cases intensifying—the very failures of traditional intermediation. The following case studies illustrate how these risks materialize in practice.

\section{Risks of Digital Advisory Platforms: Case Studies} \label{sec:casestudy}

The limitations of robo-advisors outlined in the previous section can be better understood through reference to concrete cases. This section begins with two prominent fintech platforms—Robinhood and eToro—that, while not themselves robo-advisors, demonstrate the risks that accompany digital financial intermediation. 
{Unlike robo-advisors, which typically provide automated portfolio management based on algorithmic models, Robinhood and eToro focus on self-directed trading and social investing, where users actively make decisions based on accessible platforms and social interactions.} Robinhood’s gamified interface and transaction-driven revenue model reveal how design choices intended to broaden access may also encourage impulsive trading and embed conflicts of interest. eToro, by contrast, highlights the dangers of opaque algorithmic design and social trading mechanisms that foster herd behavior and systemic fragility. 

These cases provide a foundation for examining how artificial intelligence is now being integrated into financial advisory more broadly. From conversational interfaces to advanced wealth management systems, AI applications illustrate both the opportunities of automation and the new vulnerabilities it introduces. Together, these examples underscore a central theme of this paper: that the digital transformation of financial services, without careful design and oversight, risks reproducing the same inefficiencies long associated with traditional intermediaries, while simultaneously generating new challenges.

\subsection{Robinhood and the Risks of Behavioral Manipulation}

Robinhood, a commission-free trading platform, transformed retail participation in financial markets by removing transaction barriers and offering an interface designed for ease of use. Yet its rapid expansion also raised serious concerns about how gamified design elements shaped investor behavior. The platform relied heavily on behavioral nudges—such as push notifications highlighting short-term price movements, achievement badges for trading milestones, and swipe-to-trade mechanics—that encouraged frequent activity. {During the January 2021 GameStop episode, Robinhood’s digital engagement practices—push notifications directing users to \emph{Top Movers} and \emph{100 Most Popular} lists and other gamified prompts—channeled attention toward highly volatile stocks and were explicitly deployed to stimulate trading activity and revenue \citep{Mass2020RobinhoodComplaint}. At the same time, options trading in GME surged to more than 2 million contracts on January 27, concentrated in short‑dated calls \citep{SEC2021StaffReport}. Regulators also found that Robinhood failed to exercise due diligence when approving customers for options trading, allowing many inexperienced users access to leveraged products \citep{FINRA2021RobinhoodAWC,Mass2020RobinhoodComplaint}.}

{These engagement mechanics operated alongside a business model heavily reliant on transaction‑based revenues, including payment for order flow (PFOF): the SEC sanctioned Robinhood in 2020 for misleading statements about PFOF and best execution \citep{SEC2020RobinhoodOrder}, and the company’s S‑1 discloses that a majority of revenue is transaction‑based (with \(\sim\)81\% from PFOF in Q1 2021).\footnote{Robinhood Markets, Inc., \emph{Form S‑1} (July 2021); Bloomberg Law, ``In Q1 2021 Robinhood received over \$331m in PFOF—about 81\% of revenue'' \citep{RobinhoodS1}} In combination, the attention‑shaping interface and a volume‑linked revenue model help explain why activity intensified during the episode—consistent with evidence that Robinhood users are particularly prone to attention‑induced trading \citep{Barber2022Attention}.}

The consequences extended beyond regulatory penalties. The gamified interface and incentive structure drew criticism for encouraging speculative trading, particularly among inexperienced retail investors. A tragic illustration occurred in 2020, when a young investor took his own life after misinterpreting account information related to options trading, sparking widespread public debate about the platform’s responsibility to protect its users.\footnote{\url{https://www.forbes.com/sites/sergeiklebnikov/2020/06/17/20-year-old-robinhood-customer-dies-by-suicide-after-seeing-a-730000-negative-balance/}} Robinhood has since reduced the frequency of notifications and expanded educational resources, but its reliance on transaction-driven revenues remains unchanged. The case illustrates how digital platforms can exploit behavioral biases and embed conflicts of interest, raising fundamental questions about whether their design aligns with client welfare.

\subsection{eToro — Algorithmic Rankings, Risk Communication, and Herding}

{While Robinhood highlights the dangers of behavioral design and misaligned incentives, eToro exemplifies how \emph{algorithmic opacity} can interact with investors' social behaviors to shape outcomes. The platform popularized “social trading,” allowing users to automatically replicate the trades of designated ``Popular Investors''. Visibility and copying are driven by ranking and eligibility algorithms (e.g., those that surface Popular Investors and enforce risk thresholds) and by a simplified, user‑facing risk score displayed as a 1–10 number.\footnote{eToro help center, ``Risk score explained'' (1 = extremely low; 10 = extremely high; 1–3 low, 4–6 medium, 7–10 high) and related pages on Popular‑Investor risk thresholds. \url{https://help.etoro.com/en-us/s/article/risk-score-explained-US}; \url{https://help.etoro.com/s/article/What-is-the-maximum-risk-score-for-Popular-Investors?language=en_GB}.} While marketed as tools for informed decision‑making, the precise ranking methodology is not public, and the single‑number risk presentation compresses multiple drivers (volatility, leverage, concentration) into a coarse label.}

{This paper conceptualizes "algorithmic opacity" through two distinct mechanisms. \emph{Ranking opacity} arises from non-transparent criteria that confer prominence upon certain traders, channeling attention and copy-trading activity their way. A separate issue, \emph{insufficient risk-communication}, exists when simplistic risk labels (e.g., a single 1–10 score) lack nuance and screening protocols are inadequate to restrict high-risk products to suitable clients. Both forms of opacity co-produce discrepancies between perceived and actual risk.}

{Regulatory proceedings in Australia allege exactly these mismatches for complex, leveraged products. In 2023, Australian Securities and Investments Commission
(ASIC) sued eToro, arguing that the firm’s target‑market determination (TMD) for contracts‑for‑difference (CFDs) was ``far too broad'' for a product ``where most clients lose money'', and that its screening test was ``very difficult to fail'', including instances where a client with \emph{medium} risk tolerance and no trading experience or understanding of CFD risks still fell within the target market.\footnote{ASIC, \emph{Media release 23‑204MR: ASIC sues eToro in its first design and distribution action to protect consumers from high‑risk CFD products}, 3 Aug 2023; and ASIC’s Concise Statement. \url{https://asic.gov.au/about-asic/news-centre/find-a-media-release/2023-releases/23-204mr-asic-sues-etoro-in-its-first-design-and-distribution-action-to-protect-consumers-from-high-risk-cfd-products/}; \url{https://download.asic.gov.au/media/s2qdjuat/23-204mr-asic-v-etoro-concise-statement.pdf}.} eToro’s own risk disclosures state that a majority of retail CFD accounts lose money, which demonstrates the stakes of accurate risk communication.\footnote{eToro, \emph{General Risk Disclosure} (typical disclosure that a majority of retail CFD accounts lose money). \url{https://www.etoro.com/customer-service/general-risk-disclosure/}.}}

{Beyond individual misperception, the ranking layer can concentrate retail flows in a small set of highly visible traders, generating \emph{algorithmically amplified herding}. Experimental and field evidence shows that providing salient information about ``top'' traders or enabling one‑click copying increases risk‑taking and coordination of positions: copy‑trading users follow momentum in crypto assets and display stronger co‑movement than in stocks, using eToro data \citep{kogan2024cryptos}; experimental work on social trading finds that performance signals and direct copying raise risk appetite and imitation.\footnote{CEPR VoxEU column on copy‑trading and risk taking. \url{https://cepr.org/voxeu/columns/financial-social-trading-networks-case-copy-trading-platforms}.}}

The eToro case demonstrates that AI‑mediated platform design is not immune to the risks advisory models seek to mitigate. By obscuring decision logic (ranking opacity), downplaying or simplifying risk (risk‑communication opacity), and creating feedback loops that concentrate exposures, such platforms can reproduce information asymmetries in novel forms—producing not only individual misallocation but also heightened vulnerability at the system level.

\subsection{AI Applications in Contemporary Financial Advisory}

The integration of artificial intelligence into financial advisory is not limited to the controversies observed in platforms like Robinhood and eToro. Instead, it is part of a broader trend where AI is increasingly embedded across the financial services industry. This wider adoption of AI underscores that the challenges seen in these platforms are not isolated incidents but reflect broader risks inherent in embedding AI into advisory contexts.

A prominent example is the rise of conversational agents and chatbots. Tools such as Bank of America’s ``Erica,'' Cleo in the United Kingdom {(Cleo AI Inc)}, and Allstate’s Amelia demonstrate how natural language interfaces are reshaping client engagement. By allowing clients to check balances, obtain transaction summaries, or receive savings recommendations in conversational form, these systems lower barriers to financial participation and make advisory functions more accessible. Their anthropomorphic design—presenting the system as a companion or assistant rather than a mere interface—has been shown to influence user behavior, sometimes amplifying caution or risk aversion in ways that distort financial decision-making \citep{cui2022sophia}. However, the limitations of these tools are equally significant: while they handle routine inquiries effectively, they often falter when confronted with complex or ambiguous questions, producing incomplete or inaccurate responses \citep{deriu2020spot}. The integration of generative AI has heightened these concerns, as such systems can produce fluent but misleading outputs, raising new issues of reliability, accountability, and trust \citep{ferdaus2024towards}. Privacy and data protection risks compound these challenges, given that chatbots necessarily process sensitive financial information \citep{cui2022sophia,zhu2024implementing}.

Beyond conversational interfaces, AI is being deployed in increasingly complex advisory and management functions. Wealth management platforms such as Wealthfront illustrate how multiple services—including tax optimization, liquidity management, and automated rebalancing—can be integrated into a seamless client experience \citep{balaji2024harnessing}. In the insurance sector, Oscar Health has partnered with OpenAI to apply generative AI for document summarization, claims assistance, and policy navigation, while Dai-ichi Life has developed tools such as Sophie and Ichi to support sales agents with lead generation and customer engagement. These cases highlight how AI can enhance efficiency, accelerate service delivery, and augment human advisors, particularly in high-volume or information-intensive contexts. At the same time, they raise concerns that reliance on automated systems introduces new vulnerabilities, including bias in decision-making, weak explainability, and the erosion of human reassurance in moments of financial uncertainty \citep{zhu2024implementing}.

In summary, the cases of Robinhood and eToro, alongside the broader adoption of AI, illustrate how digital advisory technologies can replicate—and, in some instances, amplify—the inefficiencies of traditional financial intermediation. Robinhood highlights the dangers of gamified interfaces and conflicted revenue models, which encourage harmful behavior. eToro demonstrates how opaque algorithms and social trading mechanisms can introduce systemic fragility. Meanwhile, the growing role of AI in both front-end client interactions and back-office operations underscores risks related to opacity, algorithmic bias, and the erosion of user trust. These risks show that while AI has the potential to revolutionize financial advisory, it must be carefully managed to avoid reproducing or exacerbating existing problems.

\section{Five Foundational Principles for Responsible AI-Driven Financial Planning}\label{sec:Prin}

Sections~\ref{sec:background} and \ref{sec:casestudy} have illustrated that while robo-advisors offer notable advantages in terms of broadening access to financial services and reducing costs, these platforms also inherit and amplify certain endemic challenges that have long characterized traditional financial intermediation. Key among these are issues such as behavioral manipulation—where design features, including gamification tactics, push notifications, and simplified trading mechanisms, encourage impulsive behavior and excessive trading; opaque recommendation engines that obscure the rationale behind portfolio advice, leaving clients unable to assess whether the recommendations genuinely align with their financial goals; and systemic herding risks that arise when multiple platforms rely on similar algorithmic models, exacerbating the risk of collective vulnerability during market crises.

Furthermore, robo-advisors are often constrained by the limitations of their own design inability to provide comprehensive, personalized, and long-term financial planning. Unlike human advisors, who can tailor advice to a client’s evolving life circumstances, tax considerations, and broader financial strategies, robo-advisors typically focus on short-term investment strategies. This gap in their advisory function underscores the challenge of relying on automated systems for complex, long-term financial needs.

These observations call for the development of a normative framework for AI-driven financial planning—one that sets clear, actionable guidelines for data usage, model development, interaction design, and governance. Such a framework must ensure that AI systems, while improving efficiency and scalability, do not compromise on transparency, accountability, or client-centricity. In particular, it is crucial that these tools operate in a manner that enhances trust and minimizes the risk of perpetuating the same structural inefficiencies observed in traditional models.

This section proposes five foundational principles—\emph{fiduciary duty}, \emph{adaptive personalization}, \emph{technical robustness and resilience}, \emph{ethical and fairness constraints}, and \emph{auditability and accountability}. The principles are complementary: fiduciary duty defines purpose, personalization ensures relevance, robustness secures stability under stress, fairness safeguards inclusion, and accountability makes responsibilities traceable and enforceable.

\subsection{Framing the Principles}\label{sec:41}

Traditional financial advice emerged to mitigate client–market information gaps and transaction frictions, yet relational asymmetries and incentive conflicts often undermined those aims. Robo-advisors promised disciplined execution at scale, but Section~\ref{sec:casestudy} shows how engagement-oriented interfaces and black-box scoring can distort behavior, obscure risk, and concentrate exposures. A principled approach is therefore needed—not to halt innovation, but to ensure that technological advances resolve rather than reconfigure the dilemmas identified earlier.

The five principles advanced here draw on two traditions. From financial services, we inherit fiduciary norms and relationship-centric logics that foreground client welfare, trust, and suitability \citep{howcroft2003banker, morgan1994commitment, vargo2004evolving, hunt2011determinants, soderberg2013relationships}. From AI, we draw on system design practices—explainability, robustness, privacy, and human-centered interaction—along with recognition of risks specific to data-driven systems such as bias, engagement-optimized objectives, and model opacity \citep{belanche2019artificial}. The aim is to translate these insights into actionable requirements for next-generation advisory platforms.

\begin{figure}[ht]
\centering
\includegraphics[width=\textwidth]{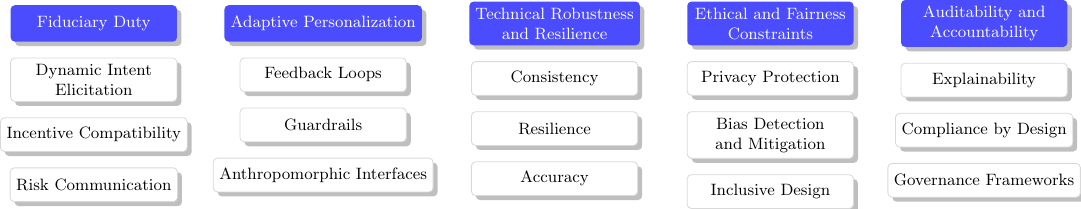}
\caption{Conceptual framework of five foundational principles for AI-driven financial planning, with subcomponents and interdependencies.}
\label{fig:principles-overview}
\end{figure}

The principles are also interdependent. Personalization without fairness can entrench exclusion; robustness without accountability can erode trust; fiduciary intent without incentive compatibility can be nullified by revenue models that privilege volume over outcomes. Figure~\ref{fig:principles-overview} provides a visual overview of the five principles and their interconnections. 

\subsection{Principle One: Fiduciary Duty—Aligning Advice with Client Interests}\label{sec:42}

The first principle sets the purpose of AI financial advice: aligning design and incentives with client welfare. In human advisory relationships, fiduciary duty is mediated by professional norms and reputation but often diluted by commissions, sales targets, or cognitive biases. AI-based advisors, in principle, need not suffer from such self-interest. In practice, however, alignment depends on how systems are built and governed. Hidden biases in training data, engagement-optimized objectives, or revenue models such as payment for order flow can reintroduce conflicts that divert advice away from client welfare \citep{belanche2019artificial}. Thus, both human and AI advisors share the fiduciary mandate, but their failure modes differ and must be addressed explicitly.

Fiduciary alignment in AI involves three design commitments. First, a \emph{comprehensive understanding of the client} is essential. Static profiling—demographics, cash flows, historical behavior—offers only a baseline \citep{ashrafuzzaman2025ai}. Effective advisors must go further, engaging in \emph{dynamic intent elicitation}: dialogue that clarifies goals, reconciles trade-offs, and translates values into operational targets. Large language models can assist by structuring such conversations, but without careful prompts and guardrails, generative systems risk literalism or overconfidence, producing plans that miss underlying purposes \citep{zhu2024implementing, inavolu2024exploring}.

{Second, fiduciary duty presupposes incentive compatibility — the alignment of provider and client interests in the advisory process. For robo-advisors and other digital platforms, this means that optimization targets (e.g., portfolio returns, engagement metrics) and monetization strategies (e.g., transaction-based fees, upselling) must not unduly prioritize provider interests at the expense of client outcomes. Common engagement proxies, such as clicks or time spent on platform, and revenue models tied to trading volume can create biases that encourage activity, rather than suitability or long-term financial planning.

To ensure compatibility between provider and client interests, several architectural responses are required. These include separating engagement metrics from financial advice objectives, limiting product offerings to those that align with client suitability and cost-effectiveness, and auditing recommendations to identify whether they favor the provider’s benefit over the client’s utility. When model opacity makes it difficult to observe whether recommendations align with fiduciary responsibility, accountability mechanisms — such as justification logs and outcome monitoring — are crucial to ensure that the system's actions can be tracked and verified to benefit the client’s interests over the long term (see Section~\ref{sec:46}).}

Finally, fiduciary alignment entails \emph{equitable treatment}. Clients with modest balances should receive the same analytical rigor as high-net-worth individuals. Because data-driven systems may inherit historical segmentation or proxy discrimination, fiduciary practice must be complemented by fairness diagnostics and governance (see Section~\ref{sec:45}). Equally important is ongoing \emph{risk awareness}: portfolios must remain consistent with clients’ capacity and willingness to bear risk, with changes communicated in ways that enhance comprehension.

In short, fiduciary duty in AI-driven financial planning is not merely a slogan but a fundamental design requirement. Without a deep understanding of the client’s needs, incentive-compatible architectures, equitable treatment, and clear risk alignment, digital platforms risk replicating the surface-level appearance of fiduciary care while failing to deliver its core substance. This framework seeks to prevent exactly those failures seen in traditional financial intermediation—such as incentive misalignment, where platforms prioritize their own profits over client welfare; opaque decision-making, where clients are left unaware of how recommendations are made or why; and behavioral manipulation, where engagement metrics and transaction-driven revenue models encourage harmful or impulsive decisions. These systemic issues are at the heart of what the fiduciary duty in AI planning seeks to address.

\subsection{Principle Two: Adaptive Personalization—Keeping Advice Relevant Over Time}\label{sec:43}

The second principle addresses relevance. Financial planning is not a one-time transaction but an evolving process shaped by shifting goals, resources, and market conditions. Traditional advisors accommodate this dynamism through periodic reviews—often annual check-ins or updates after major life events. Such episodic revision is constrained by human availability and memory. AI-driven systems, by contrast, can in principle monitor continuously, integrating behavioral signals and macroeconomic data to recalibrate advice in real time.

This capacity distinguishes AI advisors from early robo-advisors, many of which locked clients into static risk categories. A client classified as “aggressive” in an onboarding survey could remain in that allocation for years, even as their income fell, dependents increased, or market volatility spiked. During the 2018 downturn, many such investors suffered losses exceeding 30\%, revealing the dangers of rigidity. By contrast, platforms that integrate contextual cues—such as spending anomalies, income shocks, or goal milestones—can adjust allocations dynamically, making personalization an ongoing process rather than a one-off assessment \citep{wexler2021robo}.

Yet adaptivity carries its own risks. Overly reactive systems may mistake noise for signal, triggering excessive churn in portfolios and undermining long-term discipline. Algorithms that learn from behavior can inadvertently reinforce short-term biases: for instance, reducing equity exposure after temporary losses in ways that lock in underperformance. Moreover, if adjustments occur opaquely, clients may lose trust, perceiving recommendations as arbitrary. Adaptive personalization thus requires transparency and restraint: changes must be explainable, and responsiveness must be tempered by stability mechanisms (see Section~\ref{sec:44}). 

Operationalizing adaptive personalization involves two design commitments. First, systems should incorporate \emph{behavioral and contextual feedback loops}: monitoring income, expenditures, savings patterns, and engagement behaviors, while combining these with external signals such as interest-rate changes or volatility indices. Second, adaptivity should be \emph{tempered by guardrails}: hierarchical models that distinguish between temporary noise and persistent shifts, or circuit-breaker mechanisms that slow re-optimization during turbulence. These safeguards ensure that personalization serves long-term objectives rather than short-term reactions.

Equally important is how personalization is communicated. \emph{Anthropomorphic interfaces}—chatbots, voice assistants, multimodal designs—can encourage disclosure and engagement, helping capture richer data on needs and preferences \citep{adam2019investment, morana2020effect}. Yet friendliness must not be mistaken for judgment: without transparency, anthropomorphism risks fostering misplaced trust. Here, explainable AI (XAI) can clarify how observed behaviors informed updated allocations, fostering both understanding and confidence \citep{arrieta2020explainable, zhu2024implementing, belanche2020service}.

A brief illustration underscores the stakes. Consider a household that begins saving for a home down payment: a well-designed planner would reduce portfolio risk to preserve liquidity while the goal is active, then gradually restore the strategic allocation once the target is reached. By contrast, static-profile platforms can leave clients overexposed during downturns, eroding trust in automated advice. The comparison shows that adaptivity, if implemented with discipline and transparency, can transform financial planning from a static snapshot into a resilient, continuous dialogue.

In sum, adaptive personalization allows AI-driven advisors to remain relevant as clients’ lives and markets evolve. But its promise depends on design: responsiveness must be balanced with stability, behavioral learning with safeguards against bias, and continuous updates with explanations clients can understand. When those conditions hold, personalization ceases to be an onboarding exercise and becomes a durable feature of responsible financial planning.

\subsection{Principle Three: Technical Robustness and Resilience—Ensuring Stability Under Stress}\label{sec:44}

The third principle is stability. Financial advice is most valuable not in ordinary times but in moments of turbulence, when markets are volatile and client anxiety is high. Human advisors, while imperfect, can pause to reflect, apply judgment, and resist short-term panic. AI-driven advisors, by contrast, execute relentlessly according to code or learned logic. This efficiency in normal conditions becomes fragility under stress: small design flaws or data gaps can cascade into systemic failures. Robustness and resilience therefore form the backbone of responsible AI-driven planning.

The relevance of this principle lies in the technical dependencies of AI systems. Algorithms rely on continuous data flows, complex architectures, and stable infrastructure. If inputs deviate from training distributions, or if servers experience downtime, recommendations may fail silently or misfire catastrophically. Market stress amplifies these vulnerabilities: when correlations shift abruptly or volatility spikes, models extrapolating from historical patterns may deliver systematically misleading guidance. Deep learning systems are particularly prone to \emph{model drift}, where performance degrades as real-world dynamics diverge from the training environment \citep{huang2025survey, oelschlager2024evaluating}.

Robustness must therefore be designed along three dimensions. First, \emph{consistency}: systems should avoid producing contradictory advice across similar profiles or changing outputs unpredictably between sessions. Incorporating standardized decision rules and memory mechanisms helps preserve coherence. Second, \emph{resilience}: the capacity to sustain functionality during shocks, whether from adversarial inputs, sudden market dislocations, or partial data outages. Stress testing, adversarial training, and redundant infrastructure all contribute to this capacity. Third, \emph{accuracy}: predictions and optimizations must remain reliable even as conditions evolve. Continuous validation pipelines, error detection, and rollback protocols are essential to preventing compounding mistakes. These dimensions are mutually reinforcing, creating the foundation for user trust.

Balancing robustness with adaptability is crucial. Systems that prioritize stability too rigidly may fail to recognize genuine regime shifts, while systems that recalibrate too often risk overreacting to noise. The challenge is calibrated sensitivity: robust enough to withstand shocks, yet flexible enough to evolve when circumstances genuinely change. This balance connects robustness to personalization (Section~\ref{sec:43}) and, indirectly, to accountability (Section~\ref{sec:46}), since traceability is needed to evaluate whether systems adjusted appropriately under stress.

In sum, technical robustness and resilience ensure that AI-driven advisors remain trustworthy precisely when clients need them most. Without safeguards, even minor glitches can escalate into systemic risks; with them, AI systems can extend the steadiness of disciplined judgment into automated, scalable form.

\subsection{Principle Four: Ethical and Fairness Constraints—Protecting Equity in Automated Advice}\label{sec:45}

The fourth principle is fairness. Trust in financial advice depends not only on technical soundness but also on the perception that outcomes are equitable across clients \citep{hunt1994relationship}. Human advisors, though susceptible to implicit bias, operate under professional standards and legal duties that provide some guardrails. AI-driven systems, often presented as ``objective,'' risk replicating or even amplifying inequities if their data, optimization targets, or design assumptions embed hidden biases. Because financial planning directly impacts access to credit, investment opportunities, and retirement security, {failures of fairness in AI-driven systems can exacerbate existing social inequalities, leading to outcomes} that entrench disadvantage rather than alleviate it. 

The scale of algorithmic systems magnifies this concern. A biased human advisor might affect a handful of clients; a biased algorithm can affect thousands at once. Historical data used for training frequently reflect structural inequalities. If left uncorrected, models will learn and reproduce those patterns. Objectives that optimize engagement or revenue can further skew recommendations toward profitable segments, marginalizing smaller-balance investors. Adjacent evidence from algorithmic finance underscores the point: studies in credit markets show that machine learning and platform lending can yield disparate outcomes across demographic groups if fairness is not explicitly engineered \citep{bartlett2022consumer,fuster2022predictably}. These findings motivate fairness safeguards in advisory contexts as well.

Operationalizing fairness in AI-driven advice requires three commitments. First, \emph{privacy protection}: safeguarding sensitive data is a precondition for trust, particularly as advisory platforms integrate detailed personal and financial histories. Second, \emph{bias detection and mitigation}: algorithms should be routinely audited for disparate impacts, with corrective methods—such as fairness constraints in optimization, rebalancing of training samples, or outcome monitoring across demographic groups—built into governance processes. Third, \emph{inclusive design}: ensuring that clients receive comparable analytical rigor regardless of their wealth, demographics, or profitability to the provider. These commitments transform fairness from a slogan into an enforceable design principle.

Fairness is deeply interdependent with other principles. Personalization (Section~\ref{sec:43}) can improve alignment with client goals, but without fairness checks it may reinforce existing inequalities—offering sophisticated tools to wealthier clients while relegating others to generic advice. Accountability (Section~\ref{sec:46}) is also essential: without audit trails, disparate treatment remains invisible. In this sense, fairness acts as a cross-cutting constraint, shaping how all other principles are interpreted and implemented.

Practical evidence illustrates both risk and remediation. In a multi-firm review of automated investment services, the UK Financial Conduct Authority found weaknesses in firms’ suitability assessments and in how ``vulnerable consumers'' were identified and supported—raising concerns about differential outcomes across client groups and calling for process improvements to information gathering, filtering, and governance\footnote{\url{https://www.fca.org.uk/publications/multi-firm-reviews/automated-investment-services-our-expectations}}. At the same time, mission-specific platforms such as Ellevest publicly tailor portfolio design to gender-related planning horizons and risks (e.g., career interruptions, longevity), exemplifying an inclusive-design approach within legal and supervisory boundaries.\footnote{\url{https://www.ellevest.com/}} These examples indicate that fairness is not aspirational but consequential: its presence or absence produces measurable differences in long-run financial security.

In short, ethical and fairness constraints safeguard both legitimacy and inclusion in AI-driven financial planning. Without them, automation risks scaling historical inequities; with them, it can broaden access, equalize opportunity, and reinforce trust in advisory systems at scale.

\subsection{Principle Five: Auditability and Accountability—Ensuring Oversight in Algorithmic Advice}\label{sec:46}

The fifth principle is accountability. Human advisors are subject to licensing, fiduciary duties, and professional liability: when advice proves unsuitable, responsibility is traceable and sanctions are possible. By contrast, accountability in AI-driven platforms is diffuse. Responsibility may be shared among data providers, model developers, financial institutions, and regulators. When outcomes go wrong—whether due to biased data, flawed algorithms, or opaque business models—clients often face uncertainty about where liability lies. Without clear accountability, trust in algorithmic advice cannot be sustained.

Auditability is the technical foundation for accountability. Clients and regulators must be able to \emph{trace} how a recommendation was produced: what inputs were used, how they were processed, and why a given output was chosen over alternatives. This requires systematic logging of inputs, version control for algorithms, and documentation of decision rules. Auditability does not imply that every client must understand the code; rather, it ensures that independent reviewers—regulators, compliance officers, or consumer advocates—can reconstruct and evaluate system behavior when needed.

Operationalizing accountability involves three pathways. First, \emph{explainability}: advisory systems must provide either intrinsic transparency (through interpretable models) or post hoc justification (through feature attributions, natural language rationales, or visual explanations). Second, \emph{compliance by design}: legal and regulatory requirements should be embedded directly into system architecture, rather than bolted on after deployment. Third, \emph{governance frameworks}: institutions must assign explicit responsibility for monitoring, redress, and escalation, supported by independent audits and external oversight mechanisms.

The interplay with other principles is clear. Without auditability, fiduciary duty (Section~\ref{sec:42}) cannot be credibly enforced, since there is no way to verify that outputs align with client welfare. Fairness (Section~\ref{sec:45}) also depends on transparency to detect disparate treatment. Robustness (Section~\ref{sec:44}) requires traceable monitoring to catch and correct errors in real time. Accountability thus acts as the connective tissue that binds together the normative and technical requirements of trustworthy AI planning.

Illustrative cases highlight both success and failure. Charles Schwab’s Intelligent Portfolios\footnote{\url{https://www.schwab.com/intelligent-portfolios}} embed extensive documentation and compliance checks, enabling regulators to review advice logic and ensuring traceability when disputes arise. By contrast, in 2021 German regulators shut down an investment platform after discovering that its AI-driven tax optimization engine provided unsuitable advice without proper disclosures or licensing. The absence of clear accountability left clients with little recourse and forced the return of millions of euros in assets. These contrasting outcomes show that accountability is not an optional layer but a structural necessity.

In short, auditability and accountability ensure that AI-driven financial planning meets more stringent standards than human advice. {While human advisors are subject to established professional and ethical norms, AI systems must operate under heightened scrutiny due to their ability to scale rapidly and impact a large number of clients simultaneously.} By making decisions traceable, embedding compliance into system design, and assigning explicit liability, platforms can build systems that are both technologically sophisticated and institutionally trustworthy. T{his ensures that AI-driven financial tools are not only effective but also ethically sound and aligned with client interests.}

\subsection{Concluding Remarks on the Five Principles}\label{sec:47}

The five principles introduced in this section together provide a normative and operational framework for AI-driven financial planning. Each addresses a distinct vulnerability revealed in earlier sections: fiduciary alignment counters incentive distortions, personalization addresses rigidity, robustness mitigates fragility under stress, fairness prevents exclusion, and accountability ensures oversight. Taken jointly, they represent not isolated requirements but interdependent guardrails.

To further illustrate the contrasts and complementarities, Table~\ref{tab:comparison} compares traditional human intermediaries and AI-based systems across the five principles. The comparison shows that AI can surpass human advisors in scalability and adaptive personalization, but at the cost of new risks around transparency, fairness, and liability. Conversely, human advisors retain strengths in accountability and contextual judgment, though they are limited in scale and prone to subjective bias. The table underscores that the future of financial advisory is unlikely to rest on human or AI systems alone, but on hybrid arrangements that combine human oversight with algorithmic efficiency.

\begin{table}[ht]
\centering
\caption{Comparison of Traditional Human Intermediaries and AI-Based Intermediaries Under the Five Principles}
\label{tab:comparison}
\begin{tabular}{|p{4cm}|p{5.5cm}|p{5.5cm}|}
\hline
\textbf{Principle} & \textbf{Traditional Human Intermediary} & \textbf{AI-Based Intermediary} \\
\hline
Fiduciary Duty & 
- Guided by professional norms but often shaped by commissions and sales incentives. \newline
- Susceptible to subjective and unconscious bias. & 
- Can be programmed to prioritize client interests, but vulnerable to hidden incentives and data-driven distortions. \\
\hline
Adaptive Personalization & 
- Customized through periodic check-ins and interviews, but reactive and slow. \newline
- Limited ability to process large-scale data. & 
- Capable of continuous monitoring and dynamic adjustments at scale. \newline
- May risk overreacting to short-term fluctuations. \\
\hline
Technical Robustness and Resilience & 
- Vulnerable to fatigue, inconsistency, and human error. & 
- Highly efficient in stable conditions, but dependent on infrastructure and prone to systemic failures or model drift. \\
\hline
Ethical and Fairness Constraints & 
- Biases linked to individual judgment, mitigated through professional training and norms. & 
- Potential to reduce personal bias but prone to reproducing structural inequities embedded in training data. \newline
- Black-box models raise new ethical concerns. \\
\hline
Auditability and Accountability & 
- Actions are traceable and liability is clear. & 
- Decision-making often opaque and responsibility diffuse. \newline
- Requires explicit governance mechanisms for oversight and redress. \\
\hline
\end{tabular}
\end{table}

In conclusion, the five principles are not merely a checklist but the minimal conditions for legitimacy in AI-driven financial planning. They translate ethical commitments and technical best practices into operational standards, ensuring that efficiency and scalability do not come at the expense of client welfare, equity, or systemic stability. {Earlier sections identified key inefficiencies in traditional financial intermediation, such as misaligned incentives, opaque decision-making, and lack of personalization, which often undermine the quality and fairness of financial advice. The next section extends this framework by reconnecting these inherent inefficiencies to AI-driven systems, showing how—if designed with these principles—AI can transform persistent weaknesses into opportunities for more resilient, equitable, and transparent financial intermediation.

\section{A Development Roadmap for AI-Driven Financial Planning}\label{sec:roadmap}

Artificial intelligence in financial planning is not developing in isolation but along a trajectory of increasing scope and sophistication. What began as simple calculators and static tools has evolved into conversational agents, robo-advisers, and integrated planning platforms. Looking forward, the debate is not only about how advanced these systems can become but also about whether they effectively resolve the fundamental inefficiencies that have long shaped financial intermediation—information asymmetry, adverse selection, and moral hazard. 

Sections~\ref{sec:background}–\ref{sec:Prin} demonstrated that while AI can expand access, lower costs, and scale personalization, it can also reproduce entrenched vulnerabilities or introduce new ones. This motivates a structured roadmap that situates AI systems within a continuum of maturity levels, each defined not only by technical capability but also by their alignment with the five principles of responsible AI financial planning outlined in Section~\ref{sec:Prin}. 

\subsection{Introduction and Framing}  

Most existing classification schemes rank robo-advisers primarily by technical sophistication—for instance, by the complexity of algorithms or the degree of automation \citep{d2021robo, arenas2024emerging}. Such schemes are useful but incomplete: they obscure how different systems mitigate or exacerbate the economic problems that financial advice is meant to address. A chatbot with natural language fluency may appear highly advanced, yet if it offers misleading recommendations or inherits biases, it may worsen rather than resolve inefficiencies \citep{d2019promises, zhu2024implementing}. Conversely, even relatively simple tools can generate real value if they reduce information gaps or encourage disciplined behavior.  

This section therefore presents a five-level roadmap of AI-driven financial planning tools, where each stage is characterized by how well it addresses the structural dilemmas of financial advice. Rather than treating progress as a purely technological ladder, the roadmap evaluates AI maturity by two interlinked dimensions: (i) its technical and interactive sophistication, and (ii) its effectiveness in reducing market inefficiencies. The levels range from basic calculators (Level 1) to an aspirational stage of super-intelligent planners (Level 5).

To keep the discussion accessible, we introduce a concise visual in Section~\ref{sec:roadmap-overview}—an inverted‑pyramid diagram—while Appendix~\ref{appendix:metrics} provides a detailed taxonomy of metrics. These metrics are not independent dimensions; they are operational \emph{sub‑components} of the five foundational principles in Section~\ref{sec:Prin}. For example, ``dynamic intent elicitation'', ``incentive compatibility'', and ``risk communication'' operationalize fiduciary duty; ``feedback loops'' and ``guardrails'' support adaptive personalization; ``consistency'' and ``resilience'' express robustness; and ``bias detection'', ``inclusive design'', and ``privacy protection'' refine the fairness dimension.
 
Importantly, the classification metrics presented in Appendix~\ref{appendix:metrics} should not be read as independent dimensions. Rather, they represent the operational \textit{sub-components} of the five foundational principles articulated in Section~\ref{sec:Prin}. For example, ``dynamic intent elicitation'', ``incentive compatibility'', and ``risk communication'' operationalize fiduciary duty; ``feedback loops'' and ``guardrails'' support adaptive personalization; ``consistency'' and ``resilience'' express robustness; and ``bias detection'', ``inclusive design'', and ``privacy protection'' refine the fairness dimension.

\subsection{Overview of the Five Levels}\label{sec:roadmap-overview}

The five levels of AI-driven financial planning trace a progression from static tools to anticipatory, self-improving systems. Each level reflects not only advances in automation and interactivity but also how effectively the system reduces core inefficiencies. At the lower levels, AI primarily improves convenience and access; at higher levels, it aspires to holistic planning, resilience under stress, and fiduciary alignment. Figure~\ref{fig:Structure} presents the levels as a pyramid—broad at Level~1 and narrowing toward the aspirational apex at Level~5—with concise notes on core features (left) and primary limitations (right). A detailed taxonomy appears in Appendix~\ref{appendix:metrics}.

\begin{figure}[ht]
    \centering    \includegraphics[width=\textwidth]{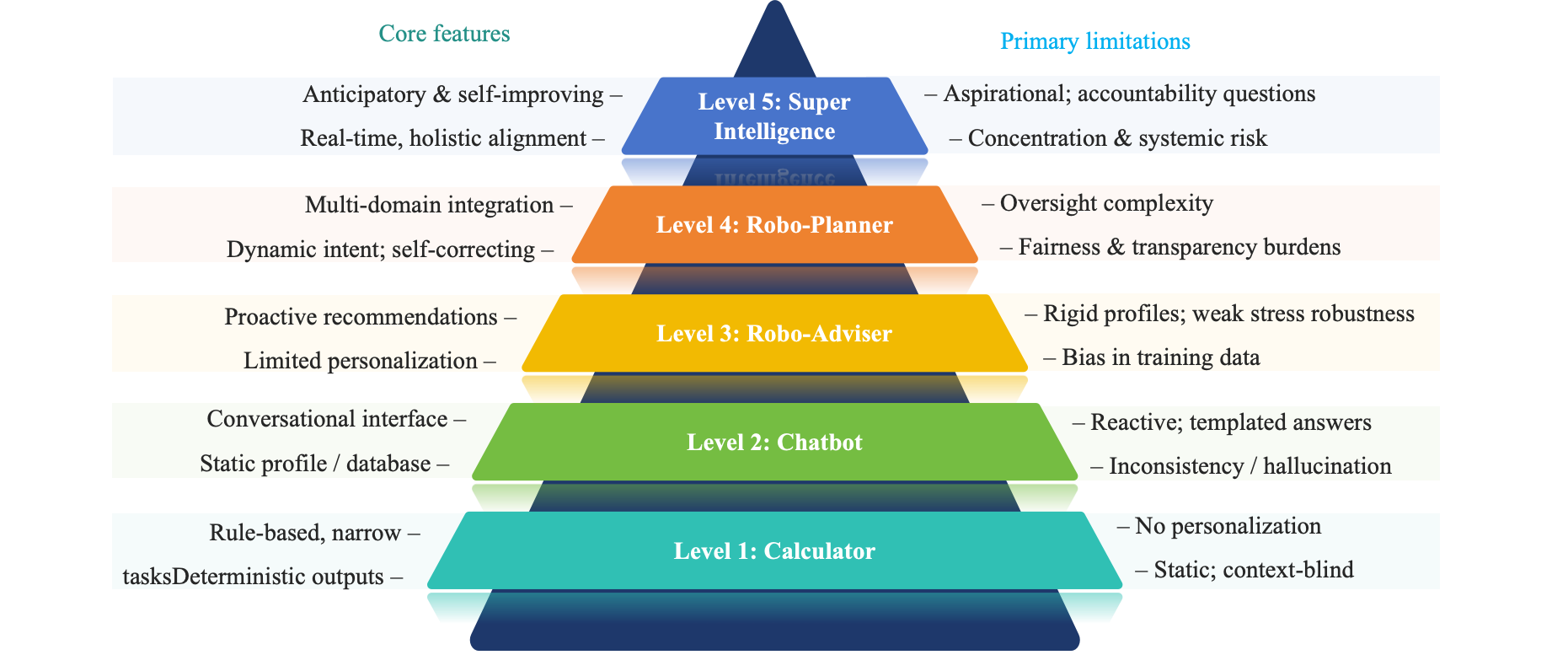}
    \caption{Pyramid roadmap of AI-driven financial planning: from foundational capabilities at Level~1 building towards the aspirational apex at Level~5. Left: \textcolor{blue}{core features}. Right: \textcolor{blue}{primary limitations}.}
    \label{fig:Structure}
\end{figure}

\subsection{Detailed Level Descriptions}\label{sec:roadmap-details}

Building on the concise overview in Figure~\ref{fig:Structure}, this subsection provides a more detailed discussion of each level. The aim is not to present a purely technological chronology, but to show how successive layers of capability interact with core financial inefficiencies such as information asymmetry, adverse selection, and moral hazard. Each level illustrates both what is gained in functionality and what remains as a limitation.

\subsubsection{Level 1: Calculator}

The first maturity level consists of deterministic, rule-based financial tools—essentially digital calculators that replicate spreadsheet functions in more accessible formats. Common examples include online mortgage calculators, retirement savings estimators, or annuity payout tables. These systems process user-provided inputs (loan amount, interest rate, time horizon) through fixed formulas to produce precise numerical outputs.

From an efficiency standpoint, calculators improve accessibility}  by making financial computations widely available to non-expert users, thereby reducing certain forms of information asymmetry. For instance, households considering a mortgage can quickly compare amortization schedules without relying on intermediaries. However, their scope is extremely narrow. Because calculators cannot contextualize client circumstances, infer intent, or adjust to external changes, they do not address adverse selection or moral hazard. Nor do they offer personalization beyond the raw parameters provided by the user.

Moreover, the determinism of calculators is a double-edged sword. While they produce consistent results for given inputs, they fail entirely outside of their pre-programmed scope. A mortgage calculator cannot account for irregular income, tax interactions, or behavioral biases that often shape real-world borrowing decisions. In this sense, Level 1 represents operational efficiency rather than advisory innovation: it removes some computational barriers but does not advance financial intermediation in a substantive sense.

Seen through the lens of the five foundational principles (Section~\ref{sec:Prin}), calculators embody almost none. They lack fiduciary alignment (they do not interpret or prioritize client interests), personalization (they cannot adapt outputs to context), robustness (they fail in edge cases), fairness (they do not address biases), and auditability (they merely display formulaic results). Their primary contribution is accessibility: enabling broader use of financial formulas that were once the preserve of professionals.

\subsubsection{Level 2: Chatbot}

The second maturity level introduces basic conversational interfaces. These systems engage users in text-based dialogue, often through natural language processing (NLP) modules, to answer questions, retrieve financial information, or simulate simple advisory interactions. Examples include early virtual assistants embedded in banking apps, capable of checking balances or suggesting basic savings accounts.

Chatbots reduce friction in accessibility by lowering the technical barrier for user interaction, thereby narrowing certain information asymmetries. They can assemble simple user profiles and deliver standardized responses in natural language, which improves usability compared to static calculators. However, their intelligence remains largely reactive: they rely on explicit user queries and struggle when goals are vague or underspecified—a frequent occurrence in financial planning. Responses are often template-driven, and results may vary with input phrasing, exposing users to logical inconsistencies or hallucinations.

In terms of financial inefficiencies, chatbots modestly improve transparency but do little to address adverse selection or moral hazard. They can surface information more conveniently but cannot interpret context, reconcile trade-offs, or align advice with long-term client welfare. Through the five-principle lens, they partially embody accessibility and, in limited form, fairness (by broadening access), but remain weak in fiduciary alignment, personalization, robustness, and accountability.

\subsubsection{Level 3: Robo-Adviser}

At Level 3, AI systems move from reactive assistance to proactive, partially personalized financial advice. Typical robo-advisers employ static profiling methods—such as risk tolerance questionnaires—to classify users as “conservative,” “moderate,” or “aggressive” investors. They then generate structured portfolio allocations or savings plans tailored to these categories. Examples include early versions of Betterment and Wealthfront, which popularized low-cost, automated investment management.

This level marks a significant step in addressing inefficiencies. By systematizing investment decisions, robo-advisers mitigate some behavioral biases (e.g., overtrading) and reduce costs relative to human advisors. They also partially reduce information asymmetry by offering retail investors access to institutional-grade diversification. Yet limitations persist. Recommendations are often locked into preset categories and cannot dynamically adjust as markets or personal circumstances change. Training data biases and opaque allocation logic can introduce hidden inefficiencies, while resilience under market stress remains weak, as seen during downturns when portfolios were not re-optimized in real time.

From the perspective of the five principles, robo-advisers represent an advance in fiduciary duty (low-cost access to diversified advice), personalization (basic tailoring through static profiles), and auditability (transparent disclosure of allocation methods). However, they remain fragile in robustness, shallow in fairness (biases often go undetected), and limited in accountability when advice proves unsuitable.

\subsubsection{Level 4: Robo-Planner}

The fourth level expands {beyond narrow investment advice to a more comprehensive model of} holistic financial planning. Robo-planners integrate multiple domains, {such as retirement planning, insurance, tax optimization, and liquidity management, into coherent, long-term strategies. For example, consider a scenario where a client uses a robo-planner to plan for retirement. The system not only suggests asset allocation but also accounts for tax implications, healthcare needs, and the client's current liquidity. It does this by continuously monitoring data such as spending patterns, income fluctuations, and market movements. Through adaptive personalization, the robo-planner adjusts strategies in real-time, dynamically revising advice based on changes in a client’s life circumstances or the broader financial environment.
} 

{To achieve this, robo-planners use dynamic intent elicitation via conversational interfaces. These interfaces help clients clarify ambiguous goals—for instance, if a client states a desire to achieve “financial independence” or “leave a legacy,” the system will guide them through a series of prompts to specify what those goals mean in financial terms (e.g., desired retirement age, wealth transfer goals). This allows the system to create more accurate financial plans that are tailored to individual preferences and life stages, rather than relying solely on predefined risk categories.

While this level of personalization is more sophisticated, it’s important to note that no current product in the market fully embodies this ideal. However, inferences can be drawn from ongoing developments in other robotic and AI-driven systems. For instance, adaptive AI systems in healthcare are beginning to integrate multiple types of data—genetic information, medical history, lifestyle factors—to provide personalized treatment plans. Similarly, AI in autonomous vehicles constantly monitors environmental factors and adjusts the vehicle’s behavior in real-time. In both cases, AI systems are moving toward more holistic, adaptive models that integrate diverse inputs to continuously improve decision-making.}

Robo-planners thus begin to address inefficiencies in financial advice more comprehensively. By synthesizing cross-domain data, they reduce issues like adverse selection (through more detailed client profiling) and moral hazard (through built-in monitoring and real-time feedback). {For example, a robo-planner might automatically flag a portfolio as ``overexposed" to a particular asset class based on a client’s evolving risk tolerance or a market shift, prompting a recalibration. Their robustness increases as they incorporate stress testing and error-detection mechanisms that simulate various market conditions to test the resilience of financial plans.
}

However, challenges remain. Opacity in decision logic—where clients cannot easily understand how recommendations are generated—could undermine accountability and trust. Complex personalization also risks reinforcing biases if fairness is not explicitly integrated into the design. For instance, if the system is trained on historical data that reflects societal inequalities, it might inadvertently favor investment options that perpetuate those disparities. Moreover, even though the business models of some robo-advisors are increasingly designed with lower fees and transparency, conflicts of interest may still persist, especially if platforms recommend proprietary products that generate additional revenues. These issues could compromise fiduciary duty, making it essential for regulators and developers to enforce safeguards.

{Viewed through the lens of the five foundational principles, Level 4 services more fully embody fiduciary alignment (via multi-domain integration), personalization (through dynamic updates), robustness (via stress-tested planning), and fairness (when paired with fairness diagnostics).} Auditability also improves through more comprehensive documentation and clear, client-facing explanations of decisions. However, the quality of implementation remains variable, and without careful governance and regulatory oversight, these gains could prove fragile, with poorly designed systems potentially undermining trust and efficacy in the long run.

\subsubsection{Level 5: Super Intelligence}

At the aspirational frontier, any super financial intelligence must be designed in alignment with the {five principles articulated in Section~\ref{sec:Prin}}. This requires three key features: its objective functions must be insulated from engagement metrics (e.g., transaction fees or upselling); a separation must exist between the platform’s business interests and those of affiliated financial institutions; external auditing and validation must also be automated to ensure machine-driven planning recommendations advance client welfare universally, without discriminating against any subgroup.{\ In practice, this alignment is evidenced by the Appendix~\ref{appendix:metrics} metrics: dynamic intent elicitation with feedback loops and guardrails (personalization), demonstrable consistency, resilience, and accuracy under stress, and explainability, compliance by design, and governance frameworks with reconstructable reasons-for-advice.}

Lessons from autonomous mobility illustrate why such alignment must be engineered, not assumed. Industry taxonomies like SAE J3016 specify \emph{levels of driving automation} but do not, by themselves, guarantee safety or accountability.\footnote{\url{https://wiki.unece.org/download/attachments/128418539/SAE\%20J3016_202104.pdf?api=v2}} High‑profile regulatory actions—such as the California DMV’s 2023 suspension of Cruise’s driverless permits following safety incidents\footnote{\url{https://www.dmv.ca.gov/portal/news-and-media/dmv-statement-on-cruise-llc-suspension/}} and the NTSB’s investigation of Uber ATG’s 2018 fatal crash in Tempe—highlight how governance, operator oversight, and system design interact under stress.\footnote{\url{https://www.ntsb.gov/investigations/accidentreports/reports/har1903.pdf}} In response, standards like UL~4600 emphasize a \emph{safety‑case} approach—comprehensive evidence that an autonomous system can operate without human supervision, with explicit provisions for fallback behavior and human override.\footnote{\url{https://users.ece.cmu.edu/~koopman/ul4600/index.html}} By analogy, a Level~5 financial planner would need a documented \emph{alignment case}: verifiable evidence that its objective functions, data pipelines, and escalation paths systematically privilege client outcomes over platform revenue. {\ This alignment case is the financial analogue of a safety case and provides concrete hooks for evaluation via our metrics (consistency/resilience/accuracy; explainability/logging).}

Healthcare AI provides a complementary analogy for model adaptivity and equity. The FDA’s AI/ML Software‑as‑a‑Medical‑Device (SaMD) Action Plan endorses a \emph{Predetermined Change Control Plan} for models that learn post‑deployment, pairing innovation with post‑market surveillance and transparency obligations.\footnote{\url{https://www.fda.gov/media/145022/download}} At the same time, empirical work shows how proxy choices can introduce structural bias: a widely used population‑health algorithm under‑referred Black patients because it optimized on \emph{cost} rather than \emph{health need}.\footnote{\url{https://pubmed.ncbi.nlm.nih.gov/31649194/}} Translated to finance, optimizing on \emph{engagement} or short‑horizon \emph{revenue} can similarly skew recommendations away from long‑term welfare unless fairness diagnostics and risk‑communication duties are built in. {\ Together, these analogies motivate the change-control and human-override guardrails we expect at Level~5.}

{Engineering challenges observed in safety-critical autonomy carry over directly to finance. Distribution shift (e.g., abrupt volatility regimes, macro or liquidity shocks) can push advisory models outside their training support, and correlated failures can emerge when many platforms implement similar logic. A Level~5 planner should therefore implement \emph{graceful degradation}: detect low-confidence or out-of-distribution inputs, slow or pause automated actions, and escalate to human review, while emitting versioned, reconstructable reasons-for-advice. In practice, this mirrors post-market change control in medical AI and the safety-case approach in autonomous mobility, and it aligns with our Appendix~\ref{appendix:metrics} metrics for guardrails, consistency/resilience/accuracy, and explainability/logging.\footnote{\url{https://www.fda.gov/medical-devices/software-medical-device-samd/artificial-intelligence-and-machine-learning-software-medical-device}}}

In principle, such systems could shrink information asymmetry (continuous, context-aware insights), alleviate adverse selection (richer screening and tailored engagement that keeps risk-averse clients invested), and curb moral hazard (pre-emptive monitoring plus transparent reasons-for-advice). They could also extend fairness via real-time diagnostics and personalization tuned to comprehension, and make accountability feasible at scale through versioned logs and explainable rationales.

Yet the very properties that make Level 5 powerful raise commensurate risks. Autonomy and scale can diffuse accountability; concentration of logic in a few platforms can create common-mode failures and systemic exposure; and over-personalization may undermine client agency. Without strict guardrails—clear liability allocation, incident reporting, independent audits, and robust override paths—capability can outpace governance. Level 5 clarifies what would be required for anticipatory planning to be both effective and legitimate, while underscoring why heightened oversight standards are essential. {\ Accordingly, the “Level~5” label is conditional on demonstrated performance against these metrics in Appendix~\ref{appendix:metrics}—not on automation alone.}

\subsection{Regulatory Considerations}\label{sec:54}

The five-level roadmap demonstrates that AI-driven financial intermediation is not a monolith but a spectrum of capabilities, each with distinct implications for long-standing market inefficiencies. At the lower levels, calculators and chatbots expand access and convenience but leave core asymmetries and incentive conflicts largely intact. Mid-level systems, notably robo-advisers and robo-planners, begin to systematize advice, reduce costs, and integrate multiple domains; yet they remain vulnerable to rigidity, bias, and fragility under stress. The aspirational apex of super-intelligent planning suggests the possibility of anticipatory, self-improving advisors, even as it raises commensurate questions about transparency, accountability, and legitimacy.

Viewed through the five principles in Section~\ref{sec:Prin}, the roadmap reveals a dual trajectory. Technical sophistication enlarges the scope for personalization and strengthens robustness, but fiduciary alignment, fairness, and accountability do not materialize automatically; they must be engineered and verified. In short, the development path of AI intermediaries is not a linear march toward more automation but an ongoing negotiation between capability and responsibility.

{The framework also has direct regulatory implications. First, advancing fiduciary duty requires incentive compatibility by design. Platforms should separate engagement proxies from the objectives that drive recommendations, disclose compensation and revenue-sharing arrangements, and apply a ``cost-dominance or justify'' rule to affiliated products—recommend them only when total cost and risk-adjusted value are superior, with a recorded rationale (cf.\ Section~\ref{sec:42}). Reasons-for-advice, fee-drag visualizations, and comprehension checks make alignment observable to clients and auditable by supervisors. Second, to ensure stability under stress (Section~\ref{sec:44}), supervisory regimes should require \emph{ex ante} stress testing of advisory logic and infrastructure, model-change controls with versioning and rollback, and graceful degradation: pausing automation and escalating to human review when inputs are anomalous or model confidence drops. Third, fairness demands routine, documented disparity testing and outcome monitoring across client groups, alongside inclusive design obligations so that advice quality and risk communication remain comparable irrespective of balance size, literacy, or channel (Section~\ref{sec:45}). Finally, accountability depends on auditability (Section~\ref{sec:46}): input and data-lineage logs, model and policy version control, incident reporting, independent audits, and clear allocation of liability among data providers, model developers, and deploying institutions.}

{These obligations should be \emph{proportionate to maturity}. For Levels~1–2, lightweight controls (transparent formulas, clear disclaimers, basic logging) suffice. At Levels~3–4, where systems make proactive, partially or fully integrated recommendations, regulators should expect comprehensive stress testing, fairness audits, rigorous risk-communication standards, and end-to-end traceability. Any movement toward Level~5 should be conditioned on a documented ``alignment case'' analogous to a safety case in other high-stakes domains: evidence that objective functions, data pipelines, and escalation paths systematically privilege client welfare, with independent assessment prior to wide deployment.}

{Taken together, the roadmap and principles yield a practical synthesis: AI can transform financial intermediation by addressing information asymmetry, adverse selection, and moral hazard, but only if capability is matched by enforceable guardrails. A principled, proportionate regime—one that couples innovation with incentive alignment, resilience, fairness, and verifiable accountability—turns the promise of automation into trustworthy advisory practice at scale.}

\section{Conclusion}\label{sec:conclusion}

The evolution of financial advice from human intermediaries to AI-driven platforms represents both a technological and an institutional transformation. This paper has shown that while digital intermediation expands access, reduces costs, and enables new forms of personalization, it also risks reproducing longstanding inefficiencies in novel ways. Case studies of Robinhood and eToro demonstrated how gamification, incentive conflicts, and opaque algorithms can distort investor behavior and amplify systemic fragility. The survey of emerging AI applications underscored that even as models become more sophisticated—through conversational agents, generative assistants, and adaptive planning—questions of transparency, fairness, and trust remain unresolved.

In response to these challenges, we articulated five foundational principles for responsible AI-driven financial planning: fiduciary duty, adaptive personalization, technical robustness and resilience, ethical and fairness constraints, and auditability and accountability. Together, these principles establish a normative framework that translates the ethical core of financial advice into actionable requirements for AI design, deployment, and oversight. They are not independent checklists but interdependent guardrails: weaknesses in one principle can undermine the effectiveness of the others.

Building on this framework, Section~\ref{sec:roadmap} introduced a developmental roadmap that categorizes AI intermediaries across five maturity levels, from basic calculators to hypothetical super-intelligent planners. The roadmap highlights both the opportunities and limitations at each stage, showing that technological sophistication alone does not ensure alignment with client welfare or systemic stability. Instead, progress must be evaluated against the principles of responsible intermediation: whether new tools genuinely mitigate information asymmetry, adverse selection, and moral hazard, or whether they simply reconfigure these inefficiencies under a different guise.

The broader implication is that the future of AI in financial planning will be determined not by technical innovation alone but by the governance structures that accompany it. Regulators, developers, and institutions must jointly ensure that AI systems remain trustworthy, inclusive, and accountable. For researchers, this calls for interdisciplinary work at the intersection of finance, computer science, and law—developing metrics to evaluate fairness, designing explainable personalization, and stress-testing algorithmic resilience under crisis conditions. For practitioners, it requires balancing efficiency with legitimacy: building systems that not only perform but also command the confidence of the clients they are meant to serve.

In sum, AI offers an unprecedented opportunity to reshape financial intermediation. Yet its promise will only be realized if efficiency and scale are matched by integrity and responsibility. The challenge ahead is to ensure that digital transformation becomes a pathway to greater financial security and equity, rather than a new frontier for hidden risks.

\begin{appendices}

\section{Extended Classification Table} \label{appendix:metrics}

Table~\ref{tab:metrics} provides the detailed taxonomy of metrics of the roadmap for AI-driven financial planning. 

\begin{table}[h!]
\tiny
\centering
\begin{tabular}{@{}p{2.5cm}p{2.2cm}p{2.2cm}p{3cm}p{3cm}p{4cm}@{}}
\toprule
\textbf{Metric} & \textbf{Level 1: Calculator} & \textbf{Level 2: Chat-bot} & \textbf{Level 3: Robo-Adviser} & \textbf{Level 4: Robo-Planner} & \textbf{Level 5: Super Intelligence} \\
\midrule
Dynamic Intent Elicitation & none & unstable, relies on explicit user queries & partial ability to infer goals in narrow domains & robust goal elicitation across domains & anticipates unspoken goals, adaptive across contexts \\
\midrule
Incentive Compatibility & none & limited, affected by interface nudges & constrained, vulnerable to training bias and opaque decisions & stronger alignment but still affected by systemic data biases & fully aligned with client welfare, transparent incentives \\
\midrule
Risk Communication & none; numeric outputs only & generic disclaimers; template warnings; no scenarios & standard risk scores and boilerplate; limited scenarios; no comprehension check & personalized drawdown/shortfall and fee-drag views; leverage/concentration sensitivity; reasons-for-advice; teach-back & real-time, context-aware explanations; counterfactuals and alerts; verified comprehension \\
\midrule
Feedback Loops & none & reacts only to simple inputs & limited behavioral feedback recognition & integrates user feedback dynamically to refine advice & continuously self-improves with behavioral and contextual feedback \\
\midrule
Guardrails & none & minimal, scripted error handling & partial safeguards (preset risk boundaries) & embedded multi-layered risk and compliance guardrails & dynamic guardrails, adaptive across regulatory and ethical domains \\
\midrule
Anthropomorphic Interfaces & none & text-based, limited persona & personalized interaction, basic social cues & rich multimodal, adaptive emotional responses & sophisticated, human-like presence with contextual adaptation \\
\midrule
Consistency & static rule-based results & variable outputs, errors under ambiguity & consistent but degrades under stress & self-corrects and maintains logical consistency & autonomously ensures reliability, evolving standards \\
\midrule
Resilience & fails outside pre-programmed inputs & fragile under volatility & robust in normal conditions, fails in extremes & stress-tested and adaptive in crises & anticipates shocks, protects portfolios dynamically \\
\midrule
Accuracy & high in narrow scope & basic factual accuracy, but hallucinations possible & solid within domain scope, but domain-limited & cross-domain accuracy with validation mechanisms & predictive foresight with near-complete accuracy \\
\midrule
Privacy Protection & none, purely local computation & basic legal compliance, vulnerable storage & standard encryption, limited guarantees & strong privacy by design, auditable protections & proactive privacy preservation, anticipates emerging risks \\
\midrule
Bias Detection and Mitigation & none & no correction & partial detection, limited mitigation & active detection and mitigation across domains & proactive bias prevention, self-adapting fairness models \\
\midrule
Inclusive Design & basic equal access; no accommodations & accessible text UI; formal compliance; little bias correction & personalized recommendations for broad users; detects and partially mitigates bias & adaptive interfaces supporting diverse needs; personalized fairness with cross-group transparency & universal inclusivity across cultures, literacy, and abilities; proactive, adaptive fairness \\
\midrule
Explainability & none & simplified summaries, little transparency & factor-based explanations & acts as tutor, explaining trade-offs and scenarios & proactive, interpretable reasoning and counterfactuals \\
\midrule
Compliance by Design & none & follows basic rules, non-dynamic & monitors compliance but lacks transparency & explainable compliance, user-auditable & self-regulates, adapts to evolving laws, predictive compliance \\
\midrule
Governance Frameworks & none & ad-hoc oversight, limited accountability & partial oversight, unclear responsibilities & embedded governance frameworks, auditable logs & autonomous governance, systemic accountability \\
\bottomrule
\end{tabular}
\caption{Updated classification metrics and levels for AI systems in financial intermediation (merging \emph{Equitable Treatment} into \emph{Inclusive Design} and adding \emph{Risk Communication} under fiduciary duty).}
\label{tab:metrics}
\end{table}

\end{appendices}

\newpage

\bibliographystyle{apalike}
\bibliography{bib.bib}

\end{document}